# Resolving the Decreased Rank Attack in RPL's IoT Networks

B. Ghaleb, A. Al-Dubai *Fellow, IEEE*, A. Hussain, *Senior Member, IEEE*, J. Ahmad, *Senior Member, IEEE,* I. I. Romdhani and Z. Jaroucheh

*Abstract*—The Routing Protocol for Low power and Lossy networks (RPL) has been developed by the Internet Engineering Task Force (IETF) standardization body to serve as a part of the 6LoWPAN (IPv6 over Low-Power Wireless Personal Area Networks) standard, a core communication technology for the Internet of Things (IoT) networks. RPL organizes its network in the form of a tree-like structure where a node is configured as the root of the tree while others integrate themselves into that structure based on their relative distance from the root. A value called the Rank is used in RPL's networks to define each node's relative position and it is used by other nodes to take their routing decisions. A malicious node can illegitimately claim a closer position to the root by advertising a lower rank value trapping other nodes to forward their traffic through that malicious node. In this study, we show how this behavior can have a detrimental side effect on the network via extensive simulations and propose a new secure objective function to prevent such an attack.

## I. Introduction

RECENTLY the Low-power and Lossy Networks (LLNs), a collection of interconnected tiny sensor nodes, have been considered one of the key enabling blocks of the ever-growing Internet of Things paradigm [1], [2]. Communication between LLNs devices is subject to restrictions on the performance of as they utilize limited resources in relation to memory footprint, processing, and power [1]. To cater for such limited resources, the Internet Engineering Task Force (IETF) has specified the IPv6 Routing Protocol for LLNs (RPL) [3] as the routing standard for such networks. Indeed, and since it was a proposal, the RPL's security aspects have been analyzed by several research efforts reporting the existence of multiple security attacks that need to be addressed to facilitate the adoption of the protocol in a wide range of applications [4][5].

RPL proposes optional cryptography modes to secure its communication aiming to provide communication integrity, confidentiality, and authenticity among other security provisions, however, the LLNs devices are not usually tamper-resistant so malicious actors can still easily get control of them and extract their security primitives to mount several types of attacks. In addition, implementing security modes of RPL can greatly degrade the network performance as many of these security primitives, such as digital signatures and encryption, are power-hungry and require an abundant of processing and storage resources that cannot be met by such resource-constrained devices [6] [7]. While some of the attacks against RPL are already well-studied in the literature as they are inherited from Wireless Sensor Networks (WSNs) such as the blackhole attack which drops received packets or delay them, some others are unique to RPL and have not yet well-studied [8] [9].

The Decreased Rank Attack is one of these attacks unique to the RPL standard. The Rank is a property in RPL which relatively represents the path quality to the Destination-Oriented Directed Acyclic Graph (DODAG) root based on routing decisions are made. Each RPL's node calculates its own rank based on a specific objective function and the rank of its preferred next hop (parent) to the root which is then communicated to immediate neighbor to calculate in turn their ranks. A node receiving multiple rank values from multiple neighbors should opt to select the neighbors with the lowest rank value as its preferred parent. Hence, the rank property can be exploited by a malicious actor, internal or external, to announce a fake lower value of the rank compared to other nodes in the network so trapping such nodes into selecting the attacker as their preferred parent towards the root. Theoretically and as reported in the literature, this attack should not be damaging as the perceived impact would be limited to building a partially sub-optimized topology where the traffic is forwarded via a fake optimal path. However, the Decreased Rank attack can be combined with other attacks to further damage the network including for instance selective forwarding or Balckhole attacks which can now be made more effective as the attacker is locating itself in a more strategic position where it receives all traffic from neighboring nodes [8][9].

In this study, the RPL Decreased Rank attack is evaluated, and a Secure Objective Function (Sec-OF) is proposed. Unlike most research studies targeting such an attack, the primary aim here is not to detect the existence of the attack or identify the attacker. It is rather to prevent a malicious actor from mounting the attack in the first place.

The rest of the paper is organized as follows. Section II briefly reviews the basic operations of RPL protocol and its security issues highlighting the rank attack. An overview of related work around the decreased rank attack is provided in Section III. The analysis of the rank attack and our proposed mechanism is presented in Section IV, followed by the performance evaluation in Section V. Finally, the conclusion and future work are reported in Section VI.

## II. BASIC RPL CONCEPTS AND OPERATIONS

RPL [3] is an IPv6 proactive distance-vector routing protocol designed by the IETF community specifically to fulfill the unique requirements of a wide range of Low-power and Lossy

Networks IoT (LLNs) applications. It organizes its physical network into a form of DODAGs where each DODAG is rooted at a single destination, referred to as the LBR (LLNs Border Router) [3][5]. The term "upward routes" is used to refer to routes that carry the traffic from normal nodes to the root (i.e., LBR) whereas routes that carry the traffic from the DODAG root to other nodes are called the downward routes [3]. The term Objective Function (OF) is used to describe the set of rules and policies that governs the process of route selection and optimization, in a way that meets the different requirements of various IoT applications [3]. In technical terms, the objective function is used for two primary goals: first, it specifies how one or more routing metrics, such as energy or latency, can be converted into a Rank, a value that reflects the node's relative position in the network; second, it defines how the Rank should be used for selecting the next hop (preferred parent) to the DODAG root. Currently, two objective functions have been standardized for RPL namely, the Objective Function Zero (OF0) [10] and the Minimum Rank with Hysteresis Objective Function (MRHOF) [11].

The OF0 is designed to select the nearest next hop to the DODAG root with no attempt to perform any load balancing. The Rank of a node is calculated by adding a strictly positive scalar value (rank-increase) to the Rank of a selected preferred parent utilizing a specific routing metric such as hop count or the expected transmission cost (ETX). For the parent selection, a node running OF0 always considers the parent with the least possible rank as its preferred parent. OF0 considers also selecting another parent as a backup in case the connectivity with its preferred parent is lost. Unlike OF0, the MRHOF is designed to prevent excessive churn (i.e., frequent parent change due to lower rank values) in the network topology and a node will not always replace change its current preferred parent to a parent with a lower rank value unless a significant change in the cost has been discovered (i.e., the Rank has changed by more than a pre-defined threshold called the Hysteresis value).

To facilitate the upward traffic pattern, a DODAG topology centered at the network root must be constructed. In such a topology, each non-root node willing to participate in upward communication must select one of its neighbors to act as that node default route (DODAG parent) towards the root [3]. The construction of the DODAG starts with the root multi-casting control messages called DODAG Information Objects (DIOs) to its RPLs neighbors. The DIOs carry the necessary routing information and configuration parameters required to build the DODAG including the rank property [3] [4]. An RPL node receiving a multicast DIO message will: (1) add the sender address to its candidate parent set; (2) calculate its distance (rank) with respect to the DODAG root based on the rank of that candidate parent, routing information advertised; (3) setup its default route (preferred parent); and (4) update the received DIO with its own rank and multicast it to other neighboring nodes, enabling them, in turn, to perform the previous operations [3][4].

### A. The Rank Decreased Attack

The RPL routing standard is vulnerable to a wide range of attacks, which can be roughly categorized into three classes [12] [13]. In the first class, the attackers aim to deplete network constrained resources such as power, bandwidth, and memory. For instance, an attack targeting the energy resources can be particularly damaging as it can greatly shorten the network lifetime and indirectly damage the network's reliability. In the second class, the attackers target the network topology usually by forcing the protocol to build sub-optimized topology or isolating some nodes from communicating with the rest of the network. In the third class, the attackers target the traffic of the network through traditional traffic analysis or eavesdropping attacks with the main aim to gather information that can help in launching the previous two classes.

The Decreased Rank attack belongs to the second class, and it is one of the most serious attacks that could mounted against the RPL protocol within the IoT 6LowPAN communication standard [8]. As mentioned earlier, the Rank property plays a crucial role in building and optimizing the routing paths in RPL's networks and under both standardised objectives functions (i.e., OF0, MRHOF), a node with a lower rank would always be preferred to take upon the next hop role towards the DODAG root. In addition to optimizing the network topology, the rank property plays a fundamental role in building a loop-free topology In the Decreased Rank attack, a malicious actor illegitimately manipulates the rank property and broadcast to its neighboring nodes a DIO (DODAG Information Object) with a fake decreased rank value. This may trigger the targeted nodes to change their preferred parents and select the attacker as their next hop to the root.

A successful attack can have a devastating impact on the network topology with major issues include: (i) non-optimized route formation, (ii) and routing loop creation. The immediate outcome of that is damaging the reliability of the network as traffic now is not forwarded through optimal routes so packet delivery ratio may be decreased, and latency is increased which is worsened by the likely formation of loops. In addition, the formation of loops would trigger RPL's repair mechanisms which requires the protocol to speed up control messages transmission (i.e., DIOs) in a useless attempt to fix the created loops. Indeed, this only has the effect of depleting network limited resources with more energy consumption and less bandwidth available for the data plane traffic exacerbating further the issue of decreased reliability and increased latency.

## III. RELATED WORK

The concepts of rank threshold and hash chain authentication have been used in [14] to mitigate several IoT attacks including the rank decreased attack carried out by an internal attacker. The proposed scheme dictates that each node should calculate its own rank, and its decrease rank threshold node upon initialization which are then hashed and advertised to the network for verification at predetermined periods. Each node is then monitored by its own parent for any deviation from expected values of the rank based on the advertised thresholds. The simulation results conducted under a simulated network is shown to have limited the impact of the attack. However, several parameters are introduced in the calculation of the threshold value requiring some fields to be added into the control messages exchanged, thus, increasing the overhead. In addition, the cryptography primitives used in the proposed may adversely affect the limited energy resources of the nodes in such networks.

A Machine Learning (ML)-based anomaly rank attack detection solution is proposed in [15] utilizing Support Vector Machines (SVMs). The developed Intrusion Detection Systems (IDS) is chosen to be deployed centrally on the border router as limited-resources normal nodes cannot tolerate the expensive operation of such a system. However, no details are provided on how the model is trained. Moreover, the tuning parameters in this study are vague and not clearly outlined.

A second ML-based rank attack detection method is developed in [17] utilizing Multi-Layer Perceptron (MLP) neural network. The operation of the proposed solution is divided into three stages. In the first stage, the rank attack is simulated using Cooja with the results saved in pcap file. The pcap file is then converted in the second stage to CSV file where the data is extracted, filtered, and converted to readable data. The MLP algorithm is applied in the third stage to detect the attack. The proposed IDS was showing to be effective in detecting the attack, however, the study aims at detecting the attack rather than to prevent it

The study in [16] proposes a new technique to detect and mitigate the impact of the rank decreased attack. The new technique utilizes two parameters to select the best parent; the rank value and a newly introduced metric called the path metric which seems to reflect a node's position in terms of number of hops. Once a node announces a new decreased rank, the new method compares it to the multiplication of a pre-determined value (e.g., 10) and the path metric and if they are not found similar, the message is considered malicious. A noticeable issue with this new method is that a non-malicious node could be classified as malicious.

A security mechanism for detecting the Rank attack is introduced in [18] based on the threshold concept. ThRankmin and ThRankmax (Minimum and Maximum Ranks thresholds) are calculated by the proposed mechanism based on the values of ranks advertised by neighboring nodes. For instance, ThRankmin is calculated by taking the mean of neighboring ranks and subtracting a fixed value (a value between 0 and 1 multiplied by the mean) from that mean. The authors chose to set the fixed value that is multiplied by the mean. While this a technique that aims to prevent the occurrence of the attack, it might be hard for a specific configuration to accurately find the optimal settings of the fixed value.

Another security mechanism called SRPL-RP is presented in [19] to detect and isolate malicious nodes that mount both the rank and version number attacks. The operation of the proposed mechanism is divided into five phases. In the first phase, timestamps are used to judge the legitimacy and the freshness of received DIOs. In the second phase, a monitoring table is created to maintain a list of parameters including the nodes' IDs upon creating the DODAG and a node is considered malicious if it is ID is not in the list. In the third and four phases, the rank values are used to detect whether a node has illegitimately decreased its rank value. The final phase is used to detect the version number attack. Results conducted by means of simulations showed that the proposed mechanism can mitigate the reported attacks. However, the approach requires timestamps to be added to the exchanged messages adding some sort of overhead.

Another framework named SVELTE is proposed in [20] for detecting routing attacks of the RPL protocol under the 6LowPAN standard (e.g., as selective-forwarding, and sinkhole). SVELTE employs a hybrid approach for intrusion detection where some modules are placed on the border router while some others are hosted on the constrained nodes of the RPL network. The framework was then evaluated by means of Contiki operating system and Cooja with a maximum number of nodes of 32. The proposed framework was shown to have a good capacity in detecting the respective attacks while not resulting in significant increase in the overhead in terms of energy consumption or memory footprint. One of the noticeable issues of SEVELTE is the unclarity regarding how to set the threshold value that governs the process of classifying nodes into malicious nodes in addition to that it only detects the existence of the attack rather than preventing it.

The authors in [21] claimed that the existing IDSs consume too much resources, thus, they developed a sink-based intrusion detection system to address the sinkhole attacks in 6LoWPAN networks. The process starts by having each node communicating to the sink some information including its IP address, preferred parent IP address, and rank encrypted with a key. The sink then compares the node's current rank to its previous rank and any node with a difference greater than a specific threshold is considered malicious. NS2 was used to evaluate the proposed and is claimed to show better detection capacity with less overhead. However, it is unclear why the messages were encrypted. In addition, the nodes are communicating a bunch of information to the sink raising the concern of significant overhead introduced especially if the network has a high churn (i.e., continues change of the preferred parent).

A hybrid anomaly-based and specification-based IDS was developed in [4] for detecting the selective forwarding and sinkhole attacks in 6LoWPAN networks. The proposed framework deploys specification-based agents in the router nodes to analyze the behavior of such nodes and send the results to the sink node. The received results at the sink node are then analyzed further using an anomaly-based agent based on the distributed MapReduce architecture to detect any malicious nodes. It was shown that the developed model achieved promising classification accuracy in comparison to other approaches in literature. However, such hybrid systems may introduce a significant overhead to the resource constrained devices and may not suit real-time systems.

## IV. THE PROPOSED SECURE OBJECTIVE FUNCTION

To address the Rank Decreased attack in RPL, a simple, but effective secure-oriented objective function has been proposed, named Secure Objective function (Sec-OF) with the basic idea is to restrict the ability of a node to change to a new parent with a better rank if that parent did not satisfy some rules. The operation time of the network is divided into two modes; normal and restricted and the network should alternate between the two modes of operations during its lifetime as follows:

1. **Normal Mode:** This is a short mode in which the network assumes that the risk of the rank attack is at its minimum, for instance, upon the initialization of the network. Hence, the protocol calculates the rank values and selects preferred parents as follows:
   - The root node (LBR) initializes the network by multicasting DIOs using Trickle timer in which it includes the routing metric to be used (e.g., initial ETX and Hop Count), and its own *Rank* alongside other network parameters.
   - A node *x* receiving the multicast DIO selects the LBR as its preferred parent *p* and proceeds into calculating its hop distance from the LBR and its own *Rank* value based on *ETX* link metric as in Eq. (1) and (2), and in turns multicasts its own DIO to its neighboring nodes.

$$h(x) = h(p) + 1 \qquad 1$$

$$Rank(x) = Rank(p) + ETX(x, p(x)) \qquad 2$$

where *h(x)* is the hop count of node *x* and *h(p)* is the hop count of the parent node which is initialized to zero for the LBR. Similarly, *Rank(x)* is the rank of node *x*, and *Rank(p)* is the rank of the parent node. *ETX (x, p(x))* is the link quality indicator between node *x* and its candidate parent *p(x)* and it represents the number of transmissions a node expects to send to a parent to successfully deliver a packet as defined in Eq. (3). In Contiki implementation of RPL, a node assigns a value between 1 and 5 to indicate the quality of the link on each packet transmission where the numbers (i.e., 1 to 5) represent how many transmissions are done before an acknowledgement is received from a neighbor. For instance, if the node receives an *ACK* from a neighbor after one transmission, the *EXT* is assigned a value of 1 for the link between that node and that neighbor and so on. An average value is then calculated, Eq. (3), for all transmitted packets using the Exponentially Weighted Moving Average (EWMA) filter, making it robust to abrupt fluctuations in link quality. Contiki RPL implementation sets the default values for the LBR *Rank* to 256 which is adopted in this study.

$$newETX(x, px) = (oldETX(x, px) * \beta + packetETX(x, px) * (Scale - \beta)) / Scale \qquad 3$$

where *oldETX(x, px)* is the current value of the *ETX*, β and *Scale* are EWMA constants, and *packetETX(x, px)* is the *ETX* of the last packet transmitted to the neighbor node *p*.
   - A node receiving DIOs from multiple other nodes should create a parent candidate set $P_k$ from its neighbors and should consider switching to a better parent when available. If it is the first time a node selects its preferred parent, it should be chosen as the parent with the minimum *Rank* as in Eq. 4. If a node has already a preferred parent *a*, and a better candidate parent *a′* in terms of Rank becomes available, then the node should switch to the new candidate parent if:

$$Rank(a') < Rank(a) - \alpha \qquad 4$$

Where α is a threshold value to reduce preferred parent switches in response to small rank changes so enhancing the stability of the network. Note that this step is similar to the parent selection method in the RPL standardised Minimum Rank with Hysteresis Objective Function (MRHOF).

2. **Restricted mode:** This is a long mode in which the network assumes that there is a higher risk of the rank decreased attack. This mode would start immediately after the normal mode, and it carries almost the same steps of the normal mode when calculating the ranks and selecting the preferred parent among a set of candidates. Specifically, the restricted mode implements Eq. (2) to calculate the rank of a node, however, it did not update the hop information. The rationale behind that is the fact that RPL networks are stationary networks in most deployments so it is highly likely that a change in the hop count of an already known candidate parent can only be a result of malicious behavior. It is worth noting that the most restrictive implementation of the proposed secure objective function would allow only for one instance of the normal mode during the lifetime of the network which is at the network initialization time. The switch from a current parent *a* to a better candidate parent *a′* in the restricted mode should only occur among nodes located at the same hop distance from the LBR and should satisfy the condition in Eq. (5).

$$Rank(a') < Rank(a) - \alpha \text{ AND } h(a') \leq h(a) \qquad 5$$

V. PERFORMANCE EVALUATION AND DISCUSSION

To evaluate the effect of the rank decreased attack on the efficiency of the network and the performance of our proposed mechanism in mitigating that attack, we have conducted a set of experiments using Contiki, a lightweight and open-source operating system designed specifically for low-power resource-constrained IoT networks [17]. Contiki features a highly optimized networking stack including several IoT standards such as Constrained Application Protocol (CoAP), UDP, 6LoWPAN and IPv6 on the top of implementing the RPL standard fundamental mechanisms. To emulate the exact binary code that runs on real sensor devices, Cooja [18], a cross-level simulator for Contiki, was used to carry out the simulation experiments. Cooja incorporates an internal hardware emulator called MSPsim [19], which is used in our simulations to impose hardware constraints of the Tmote Sky platform, an MSP430-based board with an ultra-low power IEEE 802.15.4 compliant CC2420 radio chip. We used the Unit Disk Graph Radio Medium (UDGM) radio protocol, the CSMA/CA protocol at the MAC layer and the ContikiMAC as a radio duty cycling (RDC) protocol.

The ContikiRPL library was altered to implement the Decreased Rank attack. In particular, we implemented the attack by means of a malicious node programmed to launch the attack by announcing a rank of 257 in the second minute after initializing the network. At the application layer, we simulated a periodic data collection application where each node transmits one packet to the sink every 60 seconds (the actual transmission time is randomly chosen within the 60 seconds period). We have considered in our simulations both uniform and random distribution where nodes are spread in a square area of 100m x100m. All nodes are static including the attacker and the DODAG root, which is located within the deployment area.

In the first set of experiments, we compare the performance of the two standardized objective functions (i.e., OF0 and MRHOF) in terms of reliability (Packet Delivery ratio - PDR) and power consumption with the proposed Sec-OF under various physical loss rates varying the physical link loss rate between 0% and 50%. The 0% loss rate means that the network is lossless and as a result does not experience any loss due to signal fading. However, the loss may still occur due to other factors such as hidden terminals and collisions. The simulation conditions under 0% loss will allow us to carry out experiments under near to prefect scenarios so focusing on the particulars of the evaluated protocols rather than on how they may be affected by the presence of loss. A group of 51 nodes, including one sink (Node 1), one attacker, and other 49 normal nodes is used in the first set of experiments. To investigate how the location might affect the severity of the attack, we simulated three different scenarios for the attacker in relation to the DODOAG root as follows:

- **Level 1:** we placed the attacker in the range of the DODAG root (one hop away from the root), depicted in Fig. 2 (attacker node 37).
- **Level 2:** we placed the attacker two hops away from the DODAG root, so it is in the range of at least one immediate neighbor of the root, but not in the range of the root, depicted in Fig. 3 (attacker node 7).
- **Level 3:** we placed the attacker three hops away from the DODAG root, so it is neither in the range of the root nor in the range of one of its immediate neighbours, depicted in Fig. 4 (attacker node 7).

In all scenarios, we seek to evaluate three cases: i) whether the attack is successful (i.e., nodes switched their preferred parents to the attacker node), ii) whether power consumption and PDR are affected, and iii) the extent to which performance metrics affected. **Table 1** shows a summary of the results of the first two cases under the three OFs.

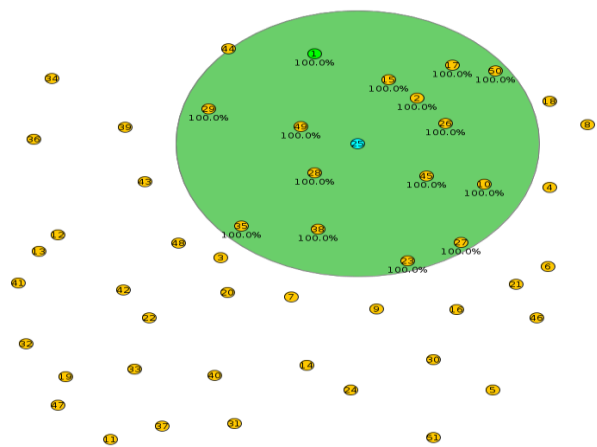

Fig. 1. Level 1: The attacker node 25 is in the immediate range of the root (node 1).

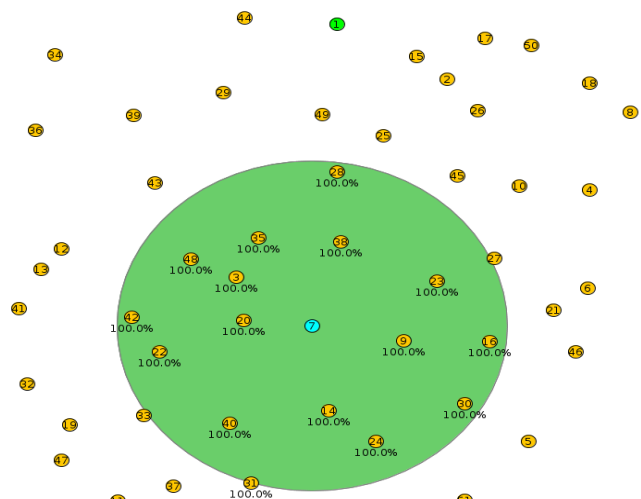

Fig. 2. Level 2: The attacker node 7 is not in the immediate range of the root (node 1), however it is in the range of some of the root neighbours.

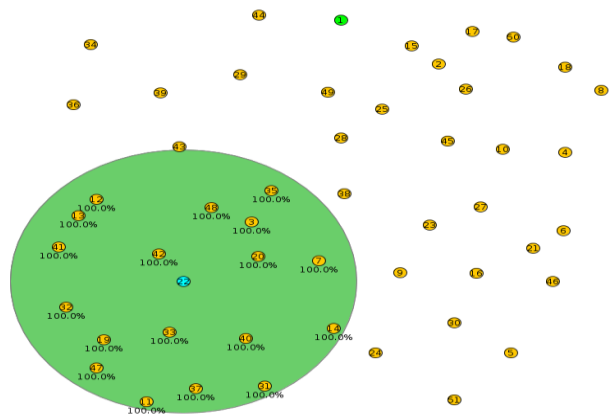

Fig. 3. Level 3: The attacker node 22 is neither in the range of the root nor in the range of one of its immediate neighbours

TABLE 1. SUMMARY OF THE RESULTS UNDER SIMULATED SCENARIOS IN A LOSSLESS NETWORK. RESULTS IN LOSSY NETWORKS ARE SIMILAR SO NOT SHOWN FOR BREVITY

| OF | Attack Successful | Power Consumption and PDR Affected |
|---|---|---|
| OF0 Level 1 | Yes, partially | NO |
| OF0 Level 2 | Yes | NO |
| OF0 Level 3 | Yes | Yes |
| MRHOF Level 1 | Yes, partially | NO |
| MRHOF Level 2 | Yes | NO |
| MRHOF Level 3 | Yes | Yes |
| OF-Sec (all levels) | NO | NO |

Table 1 shows that the attack was not successful under the proposed objective function regardless of the position of the attacker node while it was successful under both standardized OFs where the attacker is located at Level 2 or Level 3 and partially successful where the attacker is located at Level 1. The partially successful attack (some nodes selected the attacker as their parents) can be easily explained by the fact that the attacker's announced rank is greater than that of the root so only nodes which are not in the range of the root have been affected.

However, it was not obvious from looking at the general results why the attacker has affected the power consumption and the PDR when located at Level 3 but not when located at Level 1 or Level 2 under standardized OFs. Further investigation into this phenomenon has allowed us to uncover the source of the problem. In RPL networks, an extension header option "RPL Option" is used to indicate the direction of the packet using a flag named the Down 'O' flag. Hence, a packet sent by a child node to its parent should not set the Down flag indicating that the packet is heading upward and vice versa. DAG inconsistency is detected when a RPL node receives a packet with the Down 'O' bit set from a node with a higher rank (child node) and vice-versa. This case is controlled by another flag named the Rank-Error 'R' bit. When an inconsistency is detected by a node, two scenarios are possible: i) if the Rank-Error flag is not set, the forwarder node sets that flag and the packet is forwarded or, ii) if the 'R' bit is already set, the node discards the packet and the timer is reset and, DIO control messages are sent more frequently.

In fact, when the attacker is located one or two hops away from the root (level 1 and or level 2), there is no chance such an inconsistency could be marked by a forwarder node. In fact, the first forwarder node that can set the 'R' bit is the DODAG root itself which would only mark the packet as received successfully. For the level of the impact on power consumption and PDR (the third case), Fig. 5 and Fig. 6 show the effect of the attack under the evaluated OFs (Level 3) in a lossless network in terms of power consumption and PDR respectively. It is clear from the figure that under normal operations of RPL (the attack has not started) that the three OFs have comparable power consumption profiles and did not experience any packet losses and the PDR stands at around 100%. However, things started to look different after launching the attack (minute 2) where power consumption of both standardised OFs (Level 3) started to steadily increase over time and the PDR drops. The power consumption and the PDR of the proposed Sec-OF remains stable during the simulation time.

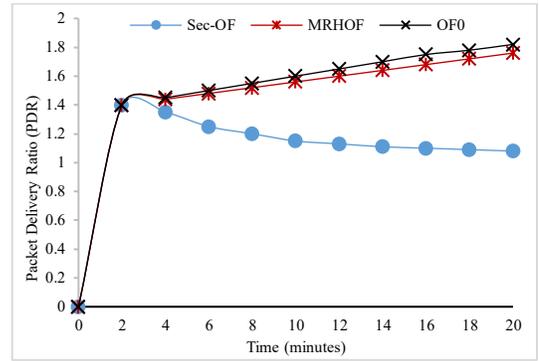

Fig. 4. The Power consumption under the three OFs (Level 3) in a lossless network. Level 1 and Level 2 attacks are not shown as they do not impact the PDR or the power consumption

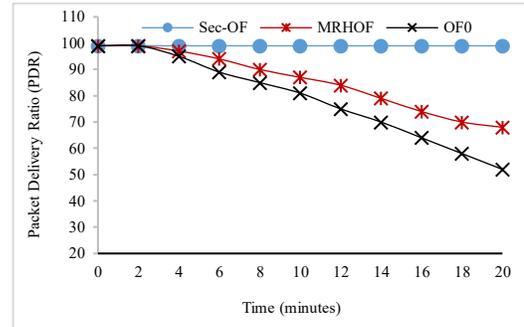

Fig. 5. The PDR under the three OFs (Level 3) in a lossless network. Level 1 and Level 2 attacks are not shown as they do not impact the PDR or the power consumption

Indeed, this was surprising to some extent as it was not expected that the attack would significantly affect the power consumption or the PDR. In fact, the Decreased Rank attack is supposed to be accompanied by other types of attacks to have an effect such as selective forwarding or blackhole attacks. Although the attack is expected to create some loops, these loops are also expected to be short-lived and the local repair of RPL should come into effect and resolve such loops. However, the reality is different and RPL under both OFs shows no capacity in resolving the created loops entering a vicious cycle that isolated the nodes in the range of the attacker from the rest of the network. This is confirmed by investigating the individual PDRs of nodes where it was found that such nodes did not manage to deliver any data-plane messages to the LBR from the moment the attack was launched. Investigating further this phenomenon, it was found that the created loops have led indirectly into a mismatch between the direction of data packets transmitted and child-parent relationships among nodes creating a case named "DODAG inconsistency". As explained earlier, in RPL, a packet should always travel upward if it is sent from a child to its parent and vice versa otherwise. This is enforced by setting some flags in the data packet transmitted and packets violating such rules are dropped according to RPL specification, a scenario that explains the unexpected data packet loss.

An interesting point is the extent to which nodes have been affected by the attack in relation to the power consumption. While only nodes in the immediate range of the attacker were affected pertinent to the PDR, the effect pertinent to the power consumption was more evident with several nodes other than those immediate neighbours showing significant higher power consumption rates. To clearly demonstrate this case, we have rerun the simulation on a smaller network composed of 11 nodes as depicted in Fig. 7 in which the attacker node has communication links with the last level of nodes (i.e., 8, 9 and 10).

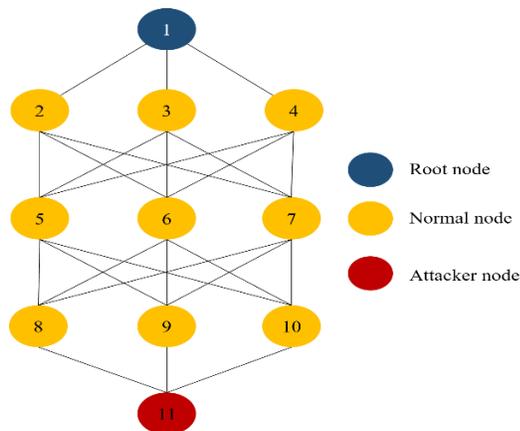

Fig. 6. A representative RPL topology used for the smaller network

Table 2 shows the individual power consumption profiles of nodes under the evaluated OFs.

TABLE 2. POWER CONSUMPTION PROFILES OF INDIVIDUAL NODES.

| Node ID | OF0 | MRHOF | Sec-OF |
|---|---|---|---|
| 2 | 1.11 | 1.12 | 1.13 |
| 3 | 1.15 | 1.06 | 1.01 |
| 4 | 1.14 | 1.05 | 1.08 |
| 5 | 1.92 | 1.57 | 1.19 |
| 6 | 1.98 | 1.73 | 1.03 |
| 7 | 2.28 | 1.71 | 1.05 |
| 8 | 2.99 | 3.25 | 0.99 |
| 9 | 3.94 | 3.01 | 1.01 |
| 10 | 4.12 | 2.85 | 0.98 |

Investigating further the source behind the unexpected poor performance of the network in terms of power consumption, it was found that three cases have contributed to that. In the first case, the nodes in the range of the attacker were found to reset their Trickle timers as a result of changing the parent node throughout the simulation time. The continues change of parent nodes indicates that the attack destabilized the network through creating loops. In the second case, the created loops forced nodes to announce a rank of infinite in an endless attempt of detaching from the DAG and then rejoin. Indeed, announcing the infinite rank to detach and to rejoin requires the nodes to reset their Trickle timers to quickly resolve such a case so further worsening the energy consumption. In the third case, it was found that the decreased rank attack has led indirectly into some sort of DAG inconsistency in relation to the data-plane traffic as explained early which contributed to both the higher energy consummation and the lower delivery ratio of the standardized OFs.

## VI. CONCLUSION

In this study, an analysis of the Decreased Rank in RPL IoT networks has been carried out. We have shown how such an attack may significantly affect the performance of the network pertaining to power consumption and PDR. The study reveals an interesting fact regarding the attack and its effect that were not reported in the literature previously. In particular, the study shows that the attack can be mounted easily without showing any effect by placing the attacker one or two hops away from the root. This is interesting as it can complicate the attack detection mechanisms that rely on network profiling so making the attack undetectable. Considering this fact, we proposed a new secure objective function, named the Secure Objective function (Sec-OF) that aims at preventing the attack from being launched in the first place. The results reported have shown the feasibility of the proposed solution in addressing the attack. It is worth noting, however, that the proposed objective function is only applicable in stationary networks and further research efforts are needed to address the decreased rank under mobile scenarios.

ACKNOWLEDGMENT

This work was supported by Edinburgh Napier University Research Starter Grants.